\documentstyle[12pt,psfig]{article}
\def\cmm2{{\,\rm cm^{-2}}}
\def\cm2{{\,{\rm cm}^2}}
\def\cmm3{{\,{\rm cm}^{-3}}}
\def\gcmm3{{\,{\rm g\,cm^{-3}}}}

\def\la{\mathrel{\mathpalette\fun <}}

\def\fun#1#2{\lower3.6pt\vbox{\baselineskip0pt\lineskip.9pt
  \ialign{$\mathsurround=0pt#1\hfil##\hfil$\crcr#2\crcr\sim\crcr}}}
\begin{document}
\pagestyle{empty}
\begin{center}
%\rightline{{\large DRAFT VERSION} (Rob 1 Mar 98)}
\bigskip
\rightline{FERMILAB--Pub--98/080-A}
\rightline{OSU-TA-98-06}
%\rightline{astro-ph/9803095}
\rightline{submitted to {\it Phys. Rev. Lett.}}

\vspace{.2in}
{\Large \bf Precision Detection of the Cosmic Neutrino
\bigskip

Background} \\

\vspace{.2in}
Robert E. Lopez,$^{1}$ Scott Dodelson,$^2$ Andrew Heckler,$^3$
and Michael S. Turner$^{1,2,4}$\\
\vspace{.2in}

{\it $^1$Department of Physics\\
The University of Chicago, Chicago, IL~~60637-1433}\\

\vspace{.1in}

{\it $^2$NASA/Fermilab Astrophysics Center\\
Fermi National Accelerator Laboratory, Batavia, IL~~60510-0500}\\

\vspace{.1in}

{\it $^3$Department of Physics\\
Ohio State University, Columbus, OH~~43210}\\

\vspace{0.1in}
{\it $^4$Department of Astronomy \& Astrophysics\\
Enrico Fermi Institute, The University of Chicago, 
Chicago, IL~~60637-1433}\\

\end{center}

\vspace{.3in}

\centerline{\bf ABSTRACT}
\medskip

\noindent  

In the standard Big Bang cosmology the canonical value for the ratio
of relic neutrinos to CMB photons is 9/11.  Within the framework of the
Standard Model of particle physics there are
small corrections, in sum about 1\%, due to slight heating of
neutrinos by electron/positron annihilations and finite-temperature
QED effects.   We show that this leads to changes in
the predicted cosmic microwave background (CMB) anisotropies
that might be detected by future satellite experiments.
NASA's MAP and ESA's PLANCK should be able to test the canonical
prediction to a precision of 1\% or better and could confirm
these corrections.

\newpage
\pagestyle{plain}
\setcounter{page}{1}

{\parindent0pt\it Introduction.}
Neutrinos are almost as abundant as photons in the Universe
and contribute almost as much energy density \cite{WEINBERG}.
Under the assumption that neutrinos decoupled completely before
electrons and positrons annihilated (at a time of around
1\,sec), the ratio of the
number density of neutrinos to that of photons is
\begin{equation}
{n_\nu \over n_\gamma} = \left( { 3N_\nu \over 11} \right)
,\label{eq:canonical}
\end{equation}
where $N_\nu = 3$ is the number of neutrino species.  Further,
because of the heating of the photons by $e^+/e^-$ annihilations,
the ratio of the neutrino temperature to the photon temperature
is $(4/11)^{1/3}=0.714$.  It follows that the ratio of the energy
density of neutrinos to that of photons is
\begin{equation}
{\rho_\nu \over \rho_\gamma} =
{7\over 8}\left( {4 \over 11}\right)^{4/3}\,N_\nu = 0.681
.\label{eq:energyratio}
\end{equation}

It has been pointed out that the assumption that
neutrinos decoupled completely before $e^+/e^-$ annihilations
is not precisely valid \cite{DICUS}.
There is now a consensus that the neutrinos share in the heating
somewhat, so
their number and energy density is slightly larger than
the canonical values, Eqs. (\ref{eq:canonical}, \ref{eq:energyratio}).
The increase is equivalent to having
slightly more than three neutrino species and
the canonical ratios.  (This is just a heuristic device, of course; the
actual number of generations is fixed at three.)
The change in the effective number of neutrino generations is
\cite{DICUS,HERRERA,RANA,DT,DOLGOV,MADSEN,DHS}
\begin{equation}
\delta N_\nu^{\rm ID} = 0.03.
\label{eq:ID}
\end{equation}
The first calculations \cite{DICUS,HERRERA,RANA}
of this effect were ``one-zone'' estimates that
evolved integrated quantities through the process of neutrino
decoupling.
More refined ``multi-zone'' calculations tracked many energy bins,
assumed Boltzmann statistics and made other approximations
\cite{DT,DOLGOV}.
The latest refinements have included these small effects as well
\cite{MADSEN,DHS}.
(A very recent calculation makes no approximations whatever and
tracks the neutrino momentum distribution over 8 orders of magnitude
in momentum \cite{GNEDIN}, and arrives at slightly higher value,
$\delta N_\nu^{\rm ID} = 0.045$.  Until the discrepancy is understood
we will stick with the earlier estimates; in any case it is simple
to rescale our results.)

There is another effect operating at roughly the same
time which acts in the same direction; it involves finite-temperature
QED corrections to the energy density of the electron, positron and
photon portion of the plasma due to interactions \cite{HECKLER,LT}.
This effect decreases the energy density of the $e^{\pm}\gamma$ plasma.
Consequently this reduces the amount of energy converted to photons when
electrons and positrons annihilate
thereby slightly raising the ratio of the neutrino to photon energy
densities.
This QED effect can also be expressed as an increase in the number
of neutrino species \cite{HECKLER,LT}
\begin{equation}
\delta N_\nu^{\rm QED} = 0.01.
\label{QED}
\end{equation}

Together, incomplete annihilation and QED finite-temperature
corrections lead to an increase in the neutrino energy density
over the canonical value by slightly more than 1\%, corresponding to
\begin{equation}
\delta N_\nu = 0.04 .
\label{dnufinal}
\end{equation}
These two corrections were initially considered in the context of Big
Bang
Nucleosynthesis.  Their net effect is to increase the predicted
$^4$He abundance by a tiny amount, $\Delta Y_P = + 5\times 10^{-4}$,
which given the present observational uncertainties, $\sigma_{Y_P}\ge
10^{-2}$, is undetectable and likely to remain so for quite
some time.  (The quantity $Y_P$ denotes the primordial mass fraction of
$^4$He.)

On the other hand, the small increase in the neutrino energy density can
have a
significant -- and potentially detectable effect -- on another
remnant of the Big Bang -- the cosmic microwave background (CMB).
In particular, the anisotropies in the CMB are very sensitive to the
epoch of matter-radiation equality, which depends on the
neutrino energy density.  As we shall show, the sensitivity
is so great that by itself the additional energy density in
neutrinos should be detectable by the very precise measurements
of the CMB anisotropy that will by made the two forthcoming
satellite experiments, NASA's MAP and ESA's PLANCK Surveyor.  However,
the situation is complicated somewhat by the fact that predictions
for the CMB anisotropies also depend on other cosmological parameters.

The aim of this paper is to address quantitatively the detectability of
the small increase in neutrino energy density
due to incomplete decoupling and finite-temperature QED effects.
For definiteness,
we assume that the primordial density perturbations that seed
structure formation and lead to CMB anisotropy
were set during inflation and allow five other parameters to vary.
They are:  baryon density ($\Omega_B$); Hubble constant ($H_0$);
amplitude of primordial perturbations; slope
of primordial perturbations ($n$); and epoch of reionization.
We find that (i) if these other parameters are held fixed
(e.g., because they are determined by other measurements), then
detectability is a sure thing; and (ii) if the other parameters
are allowed to vary, then the situation is less promising;
but, assuming non-linear effects are not a serious contaminant
and polarization of the CMB anisotropy is also measured with
precision\cite{Polarization}, these small corrections should be detectable.

{\parindent0pt \it Probing Neutrino Physics with the CMB.}
Anisotropies in the CMB are best characterized by expanding the
temperature
field on the sky in terms of spherical harmonics:
\begin{equation}
T(\theta,\phi) = \sum_{l=0}^\infty \sum_{m=-l}^l a_{lm}
Y_{lm}(\theta,\phi)
.\end{equation} 
A given theory, specified by the primordial spectrum of
perturbations and cosmological parameters, makes predictions about
the multipole amplitudes, the $a_{lm}$'s.  The predictions
take the form of statements about the distribution of the $a_{lm}$'s.
Inflationary theories typically predict that each of these coefficients
is drawn from a Gaussian distribution; as such, the distribution can
be defined by its variance.  Thus, the fundamental predictions of
inflationary models are
\begin{equation}
C_l \equiv \langle a_{lm} a^*_{lm} \rangle.
\end{equation}

Much effort has gone into computing the $C_l$'s over the last few years;
they can be calculated very accurately once the
cosmological parameters are chosen \cite{CMBCalculations}.
Viewed simplistically,
the results of a CMB experiment are estimates of the $C_l$'s,
with errors given by $\Delta C_l$.   Then, by minimizing a $\chi^2$
statistic
\begin{equation} \label{CHI}
\chi^2\left(\left\{\lambda_i\right\}\right) \equiv \sum_{l=2}^\infty 
{ \left( C_l\left(\big\{\lambda_i\big\}\right) - C_l^{\rm estimate}
\right)^2
\over (\Delta C_l)^2 }\,,
\end{equation}
the underlying set of unknown cosmological parameters
$\left\{\lambda_i\right\}$ can be estimated.

Of course, we cannot know in advance the values of $C_l$'s that
a given experiment will measure; however, by knowing what
we expect for the $\Delta C_l$'s, we can
estimate how large the uncertainties in the parameters should
be (``error forecasting'').  To do this, we
assume that the measured $C_l$'s will be
close to the true $C_l$'s.  Then, by expanding $\chi^2$
around its minimum at $\left\{ \lambda_i^{\rm true} \right\}$
we can estimate the precision to which a parameter can be
determined (for further discussion
of ``error forecasting'' in parameter estimation, see e.g.,
Refs.~\cite{PROCEDURES}):
\begin{eqnarray}\label{EXCHI}
\chi^2\left(\left\{\lambda_i\right\}\right) &&\simeq
\chi^2\left(\left\{\lambda_i^{\rm true}\right\}\right)
+ {1\over 2} \left.
{\partial^2\chi^2 \over 
\partial\lambda_i\partial\lambda_j}\right|_{\lambda=\lambda^{\rm true}} 
\left(\lambda_i - \lambda_i^{\rm true}\right) 
\left(\lambda_j - \lambda_j^{\rm true}\right)
\cr
&&\equiv \chi^2\left(\left\{\lambda_i^{\rm true}\right\}\right)
+ C_{ij} \left(\lambda_i - \lambda_i^{\rm true}\right) 
\left(\lambda_j - \lambda_j^{\rm true}\right)
. 
\end{eqnarray}
The second-derivative (Fisher) matrix $C_{ij}$
carries information about how quickly
$\chi^2$ increases as the parameters move away from their true
values.  Therefore, under some reasonable assumptions \cite{PRESS}, the
uncertainties in the parameters are determined by this matrix.
We are interested only in the 
parameter\footnote{The only other
parameter estimation paper we are aware of which considers 
$N_\nu$ as a free parameter is Jungman et al. \cite{PROCEDURES}.
Their analysis, performed several years ago, only considered $l \la
1000$ (then considered optimistic). Further they did not consider
polarization. Where it is possible to compare with them, our results
agree.}
$N_\nu$.  If all the
other cosmological parameters are held fixed,
it is a simple exercise to show that its variance is given by
\begin{equation}
\sigma_{N_\nu}^2 = {1\over C_{N_\nu ,N_\nu}}
\label{eq:novar}
\end{equation}
If all other parameters are allowed to vary, then
\begin{equation}
\sigma_{N_\nu}^2 = (C^{-1})_{N_\nu ,N_\nu}
\label{eq:var}
\end{equation}

To proceed we need to specify
\begin{itemize}

\item{} {\bf Cosmological Model.} For definiteness, we
take this to be a Cold Dark Matter model with Hubble constant $H_0 =
50\,{\rm km}\,{\rm sec}^{-1}\, {\rm Mpc}^{-1}$, baryon density
$\Omega_B = 0.08$, COBE-normalized spectrum of scale
invariant density perturbations (i.e., power-law index
$n=1$), no reionization, and energy density in cold dark matter
particles $\Omega_{\rm CDM} = 1 -\Omega_B = 0.92$.  (We assume
the simplest inflationary prediction of $\Omega = 1$.)

\item{} {\bf Experimental Errors.} Instead of tying ourselves to
a particular experiment, we assume that the experimental
uncertainty is given by
\begin{equation}
\Delta C_l = \cases{ \sqrt{2\over 2l+1} C_l & $l \le l_{\rm max}$ \cr
                        \infty & $l > l_{\rm max} $
}
. \label{eq:DCL}
\end{equation}
The error $[2/(2l+1)]^{1/2} C_l$ is the smallest possible given that
each multipole amplitude $a_{lm}$ can be sampled only $2l+1$ times;
it is the irreducible sampling or {\it cosmic}
variance.   Equation (\ref{eq:DCL}) is obviously a simplification, but
we have found it to be a reasonable approximation to the
more realistic formula \cite{KNOX} which also accounts for detector
noise.
Further, it allows us to display our results as a function of $l_{\rm
max}$,
which will give a clear sense of what angular scales need to be probed.
We use a similar formula for polarization (with different $l_{\rm
max}$).
For orientation, MAP is characterized by $l_{\rm max}
\simeq 1000$ and PLANCK by $l_{\rm max} \simeq 2500$.

\item{} {\bf Model Parameters.}  We allow
for variation in five parameters besides the neutrino energy
density: overall amplitude of the spectrum of density perturbations,
epoch of reionization parametrized by the optical depth
back to last scattering $\tau$, $H_0$, $\Omega_B$, and $n$.

\end{itemize}

\begin{figure}[tb]
\centerline{\psfig{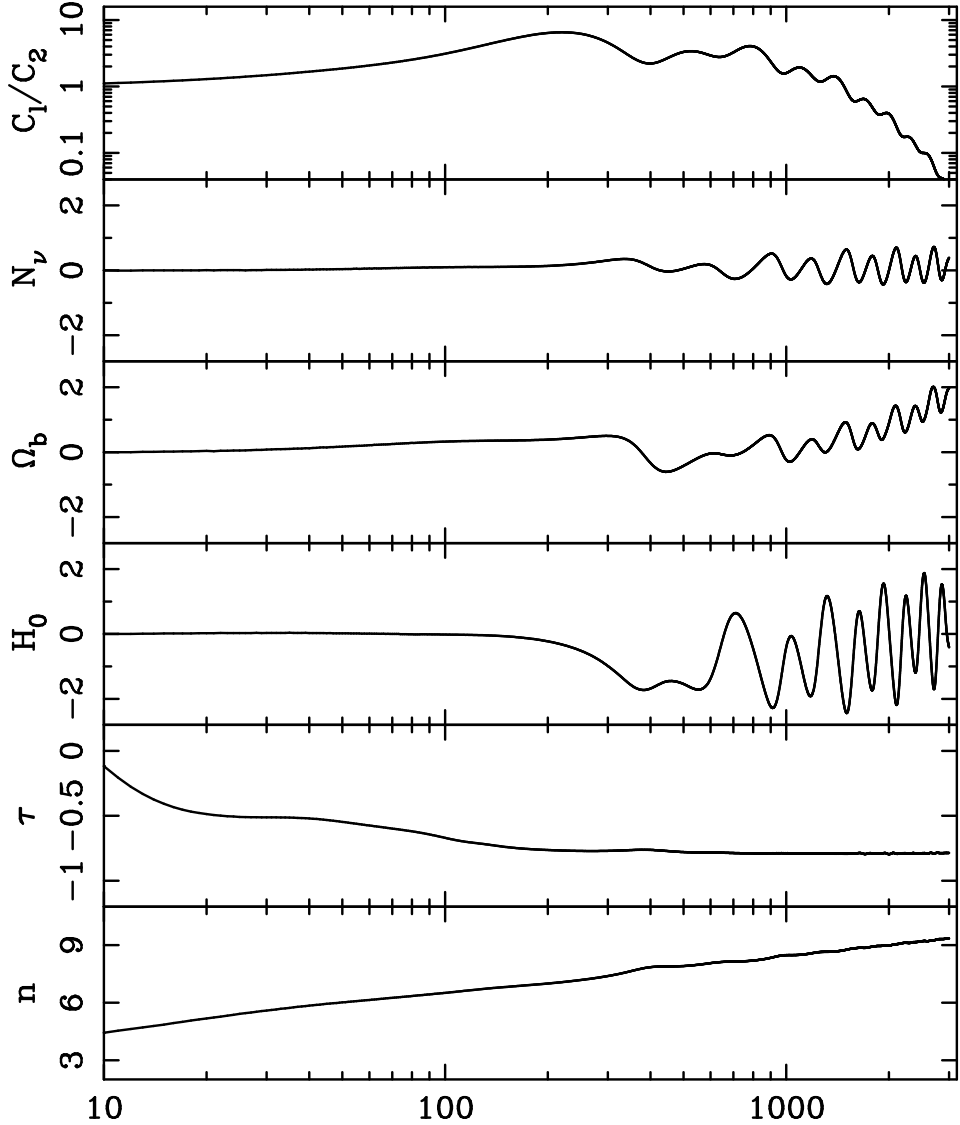}}
\caption{Temperature anisotropies and their derivatives. The top pane shows the
quadropole-normalized CMB anisotropy spectrum as a function of multipole moment $l$,
for our baseline CDM model: $N_\nu = 3$, $\Omega_B = 0.08$, $\Omega_{CDM} = 0.92$,
$H_0 = 50 {\rm km\:sec^{-1}\:Mpc^{-1}}$, scale-invariant primordial
perturbations, and no reionization. The lower panes show the derivatives of the
$C_l$'s with respect to the model parameters $\theta_i$, $\partial \ln C_l / \partial
\ln \theta_i$. For the optical depth to reionization parameter, $\partial
\ln C_l /
\partial \tau$ is plotted.}
\label{fg:fig1}
\end{figure}

\begin{figure}[tb]
\centerline{\psfig{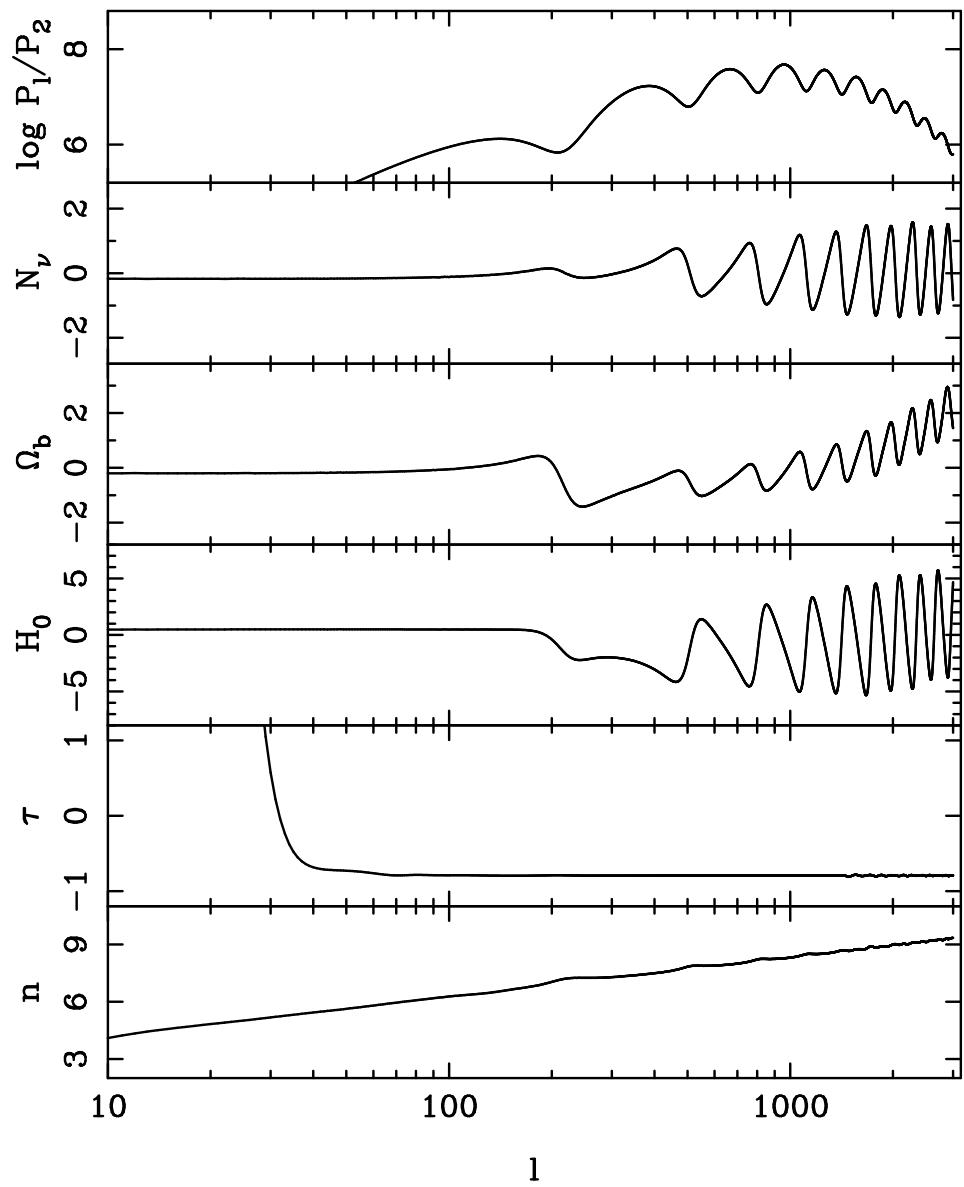}}
\caption{Same as Fig.~1, but for the $P_l$'s, the electric field polarization
anisotropies.} 
\label{fg:fig2}
\end{figure}

\begin{figure}[tb]
\centerline{\psfig{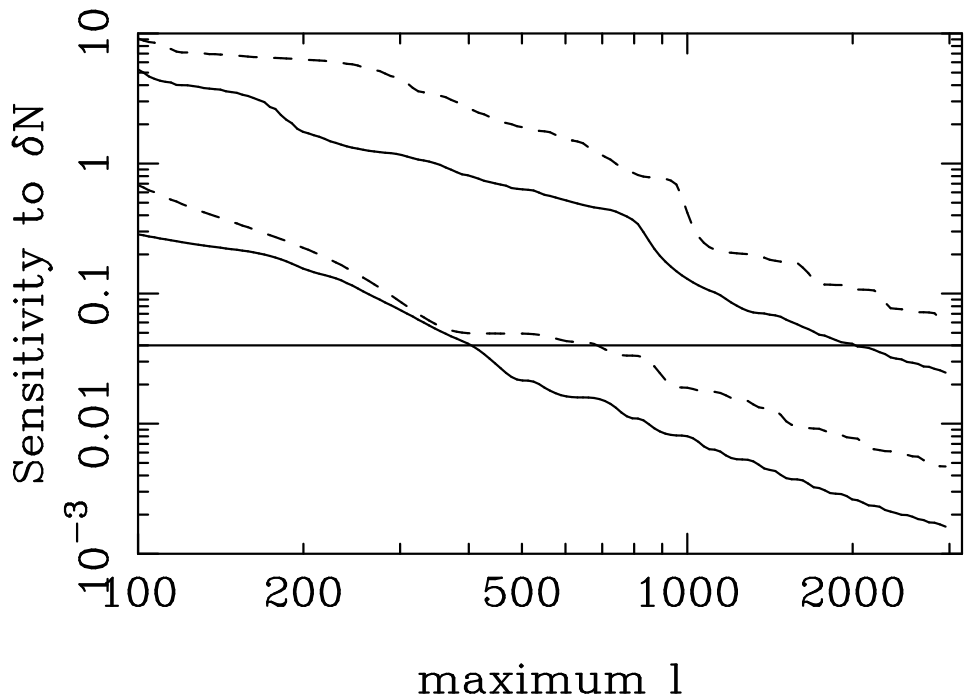}}
\caption{One-$\sigma$ sensitivity to $\delta N_\nu$, for an experiment
cosmic-variance limited up to some maximum multipole moment. The horizontal line,
$\delta N_\nu = 0.04$, is the change in effective number of neutrino families due to
neutrino heating and the QED effect. The bottom two curves are for the case where all
cosmological parameters except $N_\nu$ are fixed, while the top curves represent the
case where all parameters are determined simultaneously. For each group, the dashed
line shows the results using only temperature anisotropy data, while the solid line
shows the improvement obtained by including polarization data in the analysis.}
\label{fg:fig3}
\end{figure}

Figure~1 shows the angular power spectrum ($C_l$ vs. $l$) and how it changes as each
of the six parameters is varied.  Figure~2 shows the same for the polarization power
spectrum.  Using these derivatives, we can evaluate the Fisher matrix and
compute the expected error in $N_\nu$.
Our results are summarized in Fig.~3.

The lower set of curves in figure 3 show the
one-sigma errors on $N_\nu$ if all the other parameters
are known. Even without polarizarion information, both PLANCK
and MAP will detect the predicted $\delta N_\nu \sim 0.04$. 
If the other parameters are not well determined by other considerations,
even a very high resolution temperature anisotropy experiment will have difficulty
detecting the small predicted increase in $N_\nu$.  However, as Fig.~3 illustrates,
with polarization information, the prospects are considerably brighter.

We should note here that other effects may also mimic a change in the
effective number of neutrino species- particularly any field or
particle with a relativistic equation of state at the matter-radiation
equality epoch. For example, the presence of a
random magnetic field will contribute to the relativistic energy
density of the universe. For a given average field strength $B_{\rm
eq}$ at the radiation-matter equality epoch, one could misconstrue
this as an effective change in the number of neutrino species:
\begin{equation}
\delta N_{\nu}^{\rm Mag.} \approx 0.03\left(\frac{B_{\rm
eq}}{10\,{\rm gauss}}\right)^{2}(\Omega_0 h_{50}^2)^{-4}, 
\label{dnumag}
\end{equation}
where $h_{50}$ is the Hubble constant normalized to 50 km sec$^{-1}$
Mpc$^{-1}$. If for example we constrain $\delta N_{\nu}^{\rm Mag.}<
.01$ this translates to $B_{\rm eq}< 6\,{\rm
gauss}\,(\Omega_{0}h_{50}^2)^{2}$. Assuming that the magnetic field
$B\propto a^{-2}$, where $a$ is the scale factor, then this limit is
an order of magnitude greater than the Big Bang Nucleosynthesis limit
\cite{CHENG}. More importantly, a magnetic field this of this order
may be measured or ruled out by Faraday rotation of the CMB
\cite{KOSOWSKY} and perhaps other CMB measurements
\cite{ADAMS,BARROW}, thus the magnetic field effect may be
disentangled from $\delta N_{\nu}$.

{\parindent0pt \it Concluding Remarks.}
As our analysis shows, future, high-precision CMB anisotropy measurements have the
potential to measure the cosmic energy density in neutrinos to a precision of 1\% or
better.  Such a measurement would have significant
implications:

\begin{itemize}

\item{} If $N_\nu = 3$, further evidence for the existence
of the tau neutrino.  Note, the tau neutrino has yet to
be directly detected in the laboratory.

\item{} Determination that ``$N_\nu = 3$'' by CMB
anisotropy would confirm the canonical assumption for the
energy density in relativistic particles at the epoch of
big-bang nucleosynthesis, which is an important input parameter
for these calculations.

\item{} Confirmation of the standard
cosmology prediction that $T_\nu/T_\gamma = (4/11)^{1/3}$ to
better than 1\%.  This would test the physics of $e^+/e^-$
annihilation and neutrino decoupling in the early Universe.

\item{} Confirmation of two small physics effects
that together increase the cosmic neutrino energy density by
about 1\%.  In particular, this would be the first evidence
for finite-temperature QED corrections and a constraint to
the strength of neutrino interactions in the early Universe.

\item{} If a deviation from the expected $N_\nu = 3.04$ is found,
evidence for additional relativistic particle species (or magnetic
field) present in the early Universe or new physics in the neutrino
sector (e.g., neutrino mass or decay) \cite{DLST}.  This would have
significant implications for big-bang nucleosynthesis, structure
formation in the Universe, and elementary-particle physics.

\end{itemize}

Realizing the full potential of the CMB as a probe of the cosmic
neutrino backgrounds will require precision polarization and anisotropy
maps out to multipole number 3000.  This seems very
ambitious and perhaps even unattainable.
Nonetheless, the potential payoff discussed here makes the goal
worth striving for. If we have learned nothing
else in the years since COBE, we have certainly learned that
the experimenters have consistently manage to surprise
theorists by achieving more than was once thought reasonable.

Finally, we should acknowledge that there is room to improve
upon our analysis.  For example, we have assumed only six
cosmological parameters and ignored prior information about
them.  We have not considered the adverse role that ``secondary
anisotropies'' that are generated at late times might play.
There is clearly room for more work on this important subject.

\bigskip
\bigskip

The CMB spectra used in this work were generated by CMBFAST 
\cite{CMBCalculations}.
This work was supported by the DOE and the NASA grant NAG
5-2788 at Fermilab
and by the DoE grant DE-FG02-91ER40690 at Ohio State.

\newcommand\apj[3]{ {\it Astrophys. J.} {\bf #1}, #2 (19#3) }
\newcommand\prd[3]{ {\it Physical Review D} {\bf #1}, #2 (19#3) }
\newcommand\prl[3]{ {\it Physical Review Letters} {\bf #1}, #2 (19#3) }
\newcommand\np[3]{ {\it Nucl.~Phys.} {\bf #1}, #2 (19#3) }

\end{document}